\documentclass[mathleft
]{an}
\usepackage{graphicx}
\usepackage{times}
\begin{document}

\Pagespan{789}{}
\Yearpublication{2006}%
\Yearsubmission{2005}%
\Month{11}%
\Volume{999}%
\Issue{88}%

\title{Dynamics of the Disks of Nearby Galaxies}

\author{B. Fuchs}
\titlerunning{Dynamics of the Disks of Nearby Galaxies}
\authorrunning{B. Fuchs}
\institute{Astronomisches Rechen-Institut am Zentrum f\"ur Astronomie der
Universit\"at Heidelberg, \\  M\"onchhofstr. 12-14, 69120 Heidelberg, Germany}

\received{2008}
\accepted{2008}
\publonline{later}


\abstract{%
 I describe how the dynamics of galactic disks can be inferred by imaging and
 spectroscopy. Next I demonstrate that the decomposition of the rotation curves
 of spiral galaxies into the contributions by the various components of the
 galaxies is highly degenerate. Constraints on the decomposition can be found by
 considering implications for the dynamics of the galactic disks. An important
 diagnostic is the Toomre $Q$ stability parameter which controls the stability
 of a galactic disk against local Jeans collapse. I also show how the density
 wave theory of galactic spiral arms can be employed to constrain the mass of a
 galactic disk. Applying both diagnostics to the example of NGC\,2985 and
 discussing also the implied mass-to-light ratio I demonstrate that the inner
 parts of the galaxy, where the optical disk resides, are dominated by baryons.
 When I apply this method to the disks of low surface brightness galaxies, I
 find unexpectedly high mass-to light ratios. These could be explained by
 population synthesis models which assume a bottom heavy initial mass function
 similar to the recently proposed `integrated galactic initial mass function'.}

\maketitle

\section{Introduction}
\begin{figure}
\includegraphics[width=55mm,height=33mm]{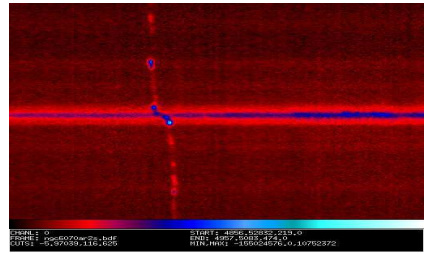}
\caption{Long slit spectrum of NGC\,6070 near the H$_\beta$ emission line. The 
dispersion of the spectrograph is oriented horizontally, while the slit 
direction points vertically.}
\label{fig1}
\end{figure}

The dynamics of the disks of spiral galaxies is inferred from observations of 
the kinematics of the stars and the interstellar gas in the galactic disks. The
most immediate way is to measure the mean rotation of the stars and the gas 
around the galactic center. Since the pioneering work of \cite{RuFo78} 
rotation curves of galaxies have been observed by long slit spectroscopy
in optical filter bands in great numbers. The rotation curves are usually 
derived using emission lines which are emitted by HII regions in the spiral 
arms of the galaxies. In Fig.~\ref{fig1} I reproduce a part of a long slit 
spectrum of NGC\,6070 close to the H$_\beta$ line (Fried \& Fuchs, unpublished).
The rotation curve -- not yet corrected for inclination -- is clearly visible to
the naked eye.
Optical rotation curves reach radially outwards typically to three radial 
exponential scale lengths of the disks.

Rotation curves observed by 21 cm emission from the interstellar HI, on the 
other hand, can usually be traced radially much further outwards. In their 
pioneering work on NGC\,3198 \cite{vanA85} could observe the HI rotation curve 
out to nearly 12 optical radial scale lengths and confirmed the existence of 
dark matter halos around spiral galaxies beyond any reasonable doubts. Rotation 
curves of both kinds are now available in large numbers and are compiled in 
catalogues like that of \cite{Pru98}.

In contrast to the mean velocities velocity dispersions are technically much 
more difficult to measure. The observed stellar absorption lines have to be 
deconvolved with the corresponding lines of template stars to measure line 
broadenings due to velocity dispersions of a few tens of km/s. Moreover spectra
at at least two slit positions along the major 
and minor axes are required to disentangle the three components of the velocity 
dispersions in the radial, tangential and vertical directions, respectively. 
This disentanglement is greatly facilitated by the fact that the ratio of radial
to tangential velocity dispersion can be related by epicyclic orbit theory 
\cite{Li59} to the shape of the rotation curves and need not to be measured 
directly\footnote{$\sigma_U/\sigma_V=2\Omega/\kappa$ to be precise.}. 
Early results were presented in the work of \cite{Bot93}. However, given the
technical possibilities of that time his results have too large errors to be
used for the purposes described below. More reliable data have been obtained by
Gerssen et al.~(1997, 2000) for the two galaxies NGC\,488 and NGC\,2985. 
Recently the {\sf disk mass} project \cite{Ver04} has begun to measure the 
velocity dispersions of face-on galaxies with the accurate {\sf Pmass}
instrument.

\section{Modelling rotation curves}
Rotation curves are modelled by constructing mass models of the observed 
galaxies. These include usually a disk with an exponential density profile 
characterized by the central surface density and a radial scale length, a bulge
model of the de Vaucouleurs or S\'ersic type characterized by the central 
density and the half-light-radius, and a model for the dark halo either with
a NFW density distribution profile \cite{Nav97},
\begin{equation}
\rho(r)=\frac{\rho_0}{r(r_{\rm c}+r)^2} ,
\label{eq1}
\end{equation}
or a quasi-isothermal sphere
\begin{equation}
\rho(r)=\frac{\rho_0}{r_{\rm c}^2+r^2} .
\label{eq2}
\end{equation}
As can be seen form eqns.~\ref{eq1} and \ref{eq2} the NFW law has a central 
density cusp, whereas the quasi-isothermal sphere has a central homogeneous 
core. Assuming essentially circular orbits of the stars and the interstellar 
gas around the galactic center - which is actually inadequate close to the 
center - the centripetal acceleration due to gravity is balanced by the 
centrifugal acceleration. For spherical systems like the bulge or the dark halo
this simply means
\begin{equation}
\frac{\upsilon_{\rm c}^2}{r}=\frac{GM(<r)}{r^2} ,
\label{eq3}
\end{equation}
with $G$ denoting the constant of gravitation, whereas for a flat exponential 
disk the circular velocity is given by \cite{Fre70} formula. The actual
rotation curve of the galaxy is then given by the sum of three terms
\begin{equation}
\upsilon_{\rm c}(r)=\sqrt{\upsilon_{\rm b}^2(r)+\upsilon_{\rm d}^2(r)
+\upsilon_{\rm h}^2(r)} ,
\label{eq4}
\end{equation}

\begin{figure}
\includegraphics[width=50mm,height=50mm]{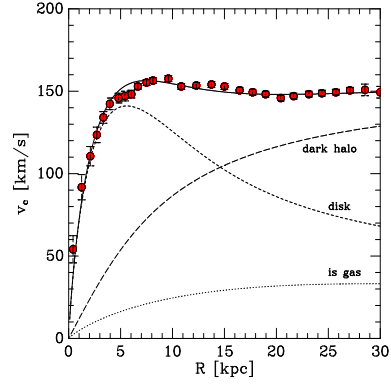}
\caption{Maximum disk decomposition of the rotation curve of NGC\,3198.
The contributions due to the stellar disk, the interstellar gas, and the dark
halo are indicated. }
\label{fig2a}
\end{figure}

\begin{figure}
\includegraphics[width=50mm,height=50mm]{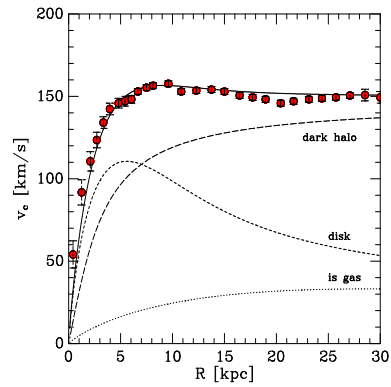}
\caption{Submaximal disk decomposition of the rotation curve of NGC\,3198. }
\label{fig2b}
\end{figure}

However, the decomposition of a rotation curve into its three components is
notoriously degenerate. I demonstrate this again with the famous example of 
NGC\,3198 which has a radial exponential scale length of $r_{\rm d}$ = 2.6 kpc.
In Fig.~\ref{fig2a} the rotation curve is fitted by a maximum 
disk model, i.e.~the disk contribution is chosen as high as allowed by the data.
Fig.~\ref{fig2b} shows a model with a submaximal disk, now of course with 
different halo parameters. Both fits cannot be distinguished by goodness-of-fit
tests from each other, but further constraints on the mass of the disk are 
needed. I have suggested for some time (Fuchs 1999) that the implied internal 
dynamical state of the stellar disks according to the galaxy models might 
provide such constraints. 
\section{Dynamical stability of galactic disks}
\begin{figure}
\includegraphics[width=83mm,height=99.6mm]{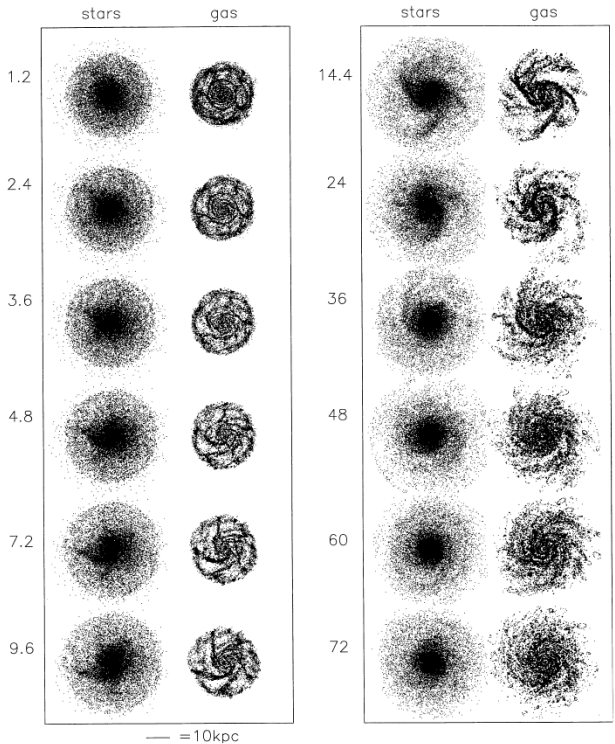}
\caption{Snapshots of a numerical simulation of the evolution of a combined 
disk of stars and gas. The disks are set up initially in radial force 
equilibrium and with Toomre parameters $Q_* = 1.5$ and $Q_g = 0.5-0.7$ 
respectively. Time is given in units of 10$^7$ yrs.}
\label{fig3}
\end{figure}

In a rotating self gravitating disk of stars there is an interplay of gravity, 
pressure, and rotation. Patches of the disk become Jeans unstable on scales 
larger than the Jeans length
\begin{equation}
\lambda \geq \lambda_{\rm J} = \frac{\sigma^2}{G\Sigma}
\label{eq5} ,
\end{equation}
where $\sigma$ denotes the velocity dispersion of the stars and $\Sigma$ is the
surface density of the disk. Detailed conservation of angular momentum sets,
on the other hand, an upper limit on the scales on which density perturbations 
can grow,
\begin{equation}
\lambda \leq \lambda_{\rm c} = \frac{G\Sigma}{\Omega^2}
\label{eq6} ,
\end{equation}
where $\Omega$ denotes the angular frequency of the patch under consideration.
If the `upper' limit is lower than the `lower' limit, 
$\lambda_{\rm c} < \lambda_{\rm J}$, growth of density perturbations is
suppressed on all scales. More precisely this
means that the \cite{Too64} stability parameter,
\begin{equation}
Q = \frac{\kappa \sigma_{\rm U}}{3.36 G \Sigma},
\label{eq7} 
\end{equation}
must be larger than 1 to ensure that a thin galactic disk is stable against 
exponentially growing density perturbations. In eq.~\ref{eq7} $\kappa$ denotes 
the epicyclic frequency of the stellar orbits,
\begin{equation}
\kappa = \sqrt{2}\, \frac{\upsilon_{\rm c}}{r} \sqrt{1+\frac{r}
{\upsilon_{\rm c}}\frac{{\rm d}\upsilon_{\rm c}}{{\rm d}r}},
\label{eq8} 
\end{equation}
and  $\sigma_U$ the radial velocity dispersions of the stars. The coefficient
3.36 applies to a Schwarzschild velocity distribution of the stars; in the case
of an isothermal gas it is replaced by $\pi$. If $Q<1$, this represents a 
serious threat to the fate of a galactic disk. In one of earliest numerical
simulations of the evolution of a self-gravitating rotating disk \cite{Ho69}
could already show that 
a thin $Q<1$ disk undergoes, even if it is rotating fast enough that the 
gravitational and centrifugal forces are in equilibrium, so violent 
instabilities that it eventually disintegrates. A $Q=1$ disk, on the other hand,
develops strong non-axisymmetric structures, but stays essentially intact. 
More recently \cite{FuLi98}
have run numerical simulations of the evolution of a combined star and gas disk.
Snapshots of one of these simulations are shown in Fig.~\ref{fig3}. Initially
the combined disk was set up in radial force equilibrium and with Toomre 
parameters $Q_*$ = 1.5 and $Q_g$ = 0.5 - 0.7. Such a disk has an effective $Q$
parameter of slightly less than 1. The disk became immediately unstable as 
witnessed by the ring like structures especially in the gas disk. The rings
fragment into clumps and shear then due to the differential rotation into
spiral arms. These grow then rapidly to large amplitudes, both in the stellar 
and gas disks. The perturbations become so strong that they heat up the stellar
disk dynamically so much that all spiral structures are eventually wiped out
in it. A disk with $Q<1$ is not an equilibrium model!

\begin{figure}
\includegraphics[width=50mm,height=50mm]{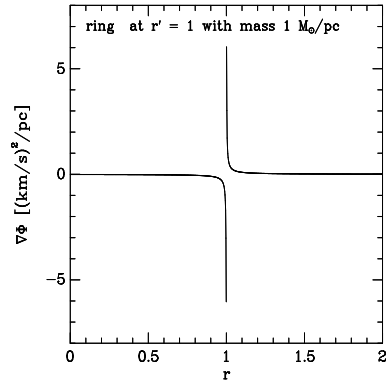}
\caption{The radial gravitational force of a mass loaded thin ring calculated 
in the plane of the ring. The forces diverge at the edges of the ring.}
\label{fig4}
\end{figure}

The physical reason that galactic disks are prone to such fierce gravitational
instabilities lies in their thin geometry. One can imagine a disk as an 
assembly of thin rings. The gravitational attraction does not vanish inside the 
ring like in a spherical  shell, but diverges to infinite repulsion at the inner
edge. At the outside edge of the ring the gravitational attraction becomes 
infinite regardless of the actual surface density of the ring. In
Fig.~\ref{fig4} I illustrate an example.
An assembly of such mass loaded rings must be obviously a highly unstable 
system.
\begin{figure}
\includegraphics[width=65mm,height=33mm]{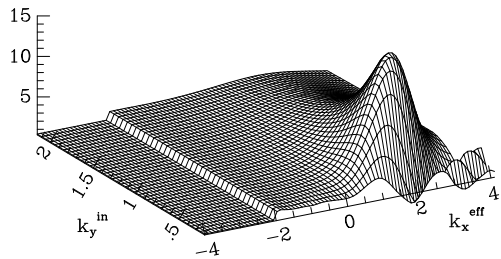}
\caption{Amplification factor of the swing-amplification mechanism for spiral 
density waves. The amplification factor is shown as function of the 
circumferential wave number $k_y$ and the radial wave number $k_x$, both
measured in units of $k_{\rm crit}$, assuming a flat rotation curve,
$ A=\Omega/2$. Peak amplification occurs at $k_y = 0.5\, k_{\rm crit}$.}
\label{fig5}
\end{figure}

\section{Density wave theory of spiral arms.}
The density wave theory predicts the preferential wave length $\lambda$ of 
spiral density waves. I adopt here the concept of swing-amplification theory
(Toomre 1981) which predicts
\begin{equation}
\lambda = {\mathcal{X}}(A/\Omega)\, \lambda_{\rm crit}
\label{eq9} 
\end{equation}
with the natural length unit
\begin{equation}
\lambda_{\rm crit}=\frac{4\,\pi^2G\,\Sigma}{\kappa^2} .
\label{eq10} 
\end{equation}
The coefficient ${\mathcal{X}}$ in eq.~\ref{eq9} depends on the slope of the 
rotation curve measured by Oort's constant $\frac{A}{\Omega}=
-\frac{1}{2}\frac{r}{\Omega}\frac{{\rm d}\Omega}{{\rm d}r}$. In Fig.~\ref{fig5}
I reproduce 
a figure of the operation characteristics of the swing-amplification mechanism 
taken from \cite{Fu01}. The amplification factor is shown as 
function of the circumferential wave number $k_{\rm y}=2\,\pi/\lambda$ and the
radial wave number $k_{\rm x}$ assuming a flat rotation curve, $ A=\Omega/2$.
As can be seen amplification becomes maximal at $k_{\rm x}$ around 2 and 3
$k_{\rm crit}$( $k_{\rm crit} = 2\,\pi/\lambda_{\rm crit}$). Peak amplification
occurs in this example of a flat rotation curve at $k_{\rm y}=0.5$ or
$\lambda = 2 \lambda_{\rm crit}$. A relation for the coefficient
${\mathcal{X}}(A/\Omega)$ for other slopes of the rotation curve is given 
in \cite{Fu01}.

\begin{figure}
\includegraphics[width=50mm,height=32mm]{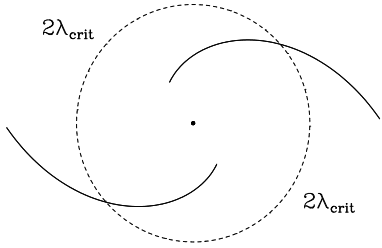}
\caption{Sketch of how the preferred wave length -- $2\,\lambda_{crit}$ in this 
example -- is related to the number of spiral arms.}
\label{fig6}
\end{figure}

\cite{Ath87} have pointed out that eq.~\ref{eq9} can be used to constrain the
mass of the disk of a spiral galaxy. In Fig.~\ref{fig6} I illustrate with a 
sketch how the circumferential wave length is related to the number of spiral 
arms $m$. Obviously the number of arms is determined by how often the
preferred wave length fits onto the annulus,
\begin{equation}
m=\frac{2\,\pi\, r}{\lambda}=\frac{2\,\pi\, r}{{\mathcal{X}}
\lambda_{\rm crit}} .
\label{eq11} 
\end{equation}
Thus one can infer from the number of spiral arms $m$, if the rotation curve is
known, the critical wave length $\lambda_{\rm crit}$, and from thence the 
surface density $\Sigma$ (cf.~eq.~\ref{eq10}). 
This concept has been tested successfully 
with numerical simulations of the spiral structure of galactic disks. Since the
parameters of the disk which have been adopted for the simulation are known,
one can read eq.~\ref{eq11} from right to left and compare the predicted and 
actually found number of spiral arms. For instance, in the simulation by 
\cite{FuLi98} the predicted number of spiral arms is $m$ = 3 which can be 
clearly confirmed in the snapshot at t = 14.4$\cdot$10$^7$ yrs in 
Fig.~\ref{fig3}.

Often it is argued that spiral density waves do not grow spontaneously in the 
disks of galaxies, but are induced by tidal forces during encounters of two 
galaxies. Such events do occur, but are not long lived. Zang \& Toomre 
(Toomre 1981) have demonstrated with their work on the Mestel disk that such 
induced spiral density waves wrap up and die out on the timescale of one 
galactic revolution. Similarly \cite{Dub08} have shown that in their numerical
simulations satellite galaxies orbiting around galaxies induce in the galactic
disks of the parent galaxies transient, swing amplified density waves.

\section{Example: Diagnostics of NGC\,2985}
\begin{figure}
\includegraphics[width=60mm,height=44mm]{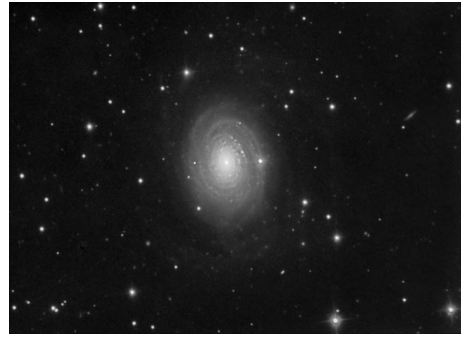}
\caption{Image of NGC\,2985. (Source: DSS)}
\label{fig7}
\end{figure}

\begin{figure}
\includegraphics[width=50mm,height=50mm]{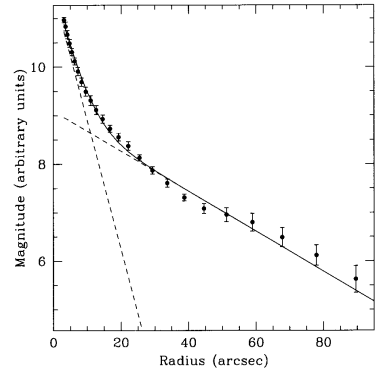}
\caption{Surface density profile of NGC 2985. The dashed lines are fits of 
the bulge and the exponential disk, respectively. The solid line is the
superposition of both components.}
\label{fig8}
\end{figure}

\begin{figure}
\includegraphics[width=65mm,height=60mm]{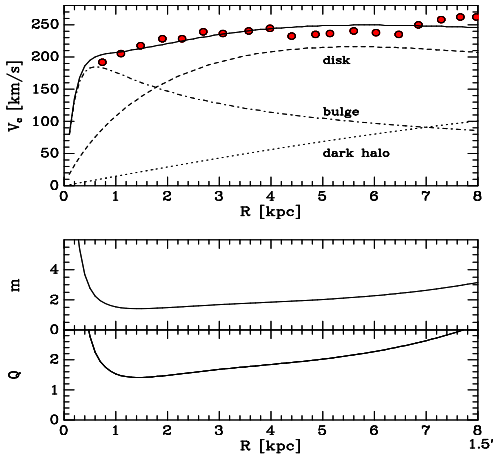}
\caption{Upper panel: Maximum disk model of the rotation curve of NGC\,2985. 
The contributions by the various components are indicated as dashed lines. 
Lower panel: Predicted number of spiral arms $m$ and the Toomre $Q$ parameter. 
On the abscissa the scale is also indicated in arcminutes; 
1.5 arcmin corresponds to the optical radius of the disk.}
\label{fig9}
\end{figure}

NGC 2985 is a nearby (D = 18 Mpc) bright spiral galaxy. An image taken from DSS,
which is reproduced in Fig.~\ref{fig7}, shows that it has developed well 
defined bisymmetric spiral arms.
\cite{Ge00} have measured its rotation curve and the velocity dispersions of 
the stars in two slit positions so that the three dimensional velocity 
dispersion components are available. Moreover, they have imaged the galaxy and 
derived its surface density profile which is shown in Fig.~\ref{fig8}. The 
authors have also provided a bulge-disk model for the photometry by fitting a 
bulge with an exponential density profile and an exponential disk with a radial
scale length of $r_{\rm d}$ = 2.6 kpc to their data.
\cite{ArFu01} have changed this model slightly. They modelled the bulge by a
softened power law, $\rho \sim (r_{\rm c}^2+r^2)^{-1.75}$ with $r_{\rm c}$
= 275 pc, added a dark halo component and calculated rotation curves. In
Fig.~\ref{fig9} I show first a maximum disk model 
in which the scale radii and the bulge-to-disk ratio
as found in the photometric model are adopted, but the mass-to-light ratio is
chosen as high as allowed by the kinematical data. As can be seen in 
Fig.~\ref{fig9} hardly any dark matter is required to fit the observed 
rotation curve. In the lower panel I show the predicted number of spiral arms 
and the $Q$ parameter. Both diagnostics are consistent with the fact that the
dynamical state of the disk of NGC\,2985 allows the galaxy to develop a 
two-armed spiral. The implied mass-to-light ratio is $M/L_B$ = 2.0
$M_\odot/L_\odot$. This is in good agreement with the prediction of 
population synthesis models. NGC\,2985 has a $B-R$ colour of $B-R$ = 1.1 mag.
For such a colour the population synthesis models compiled by \cite{BeJo01}
predict a mass-to-light ratio of about $M/L_B$ = 2.8 $M_\odot/L_\odot$.

\begin{figure}
\includegraphics[width=65mm,height=60mm]{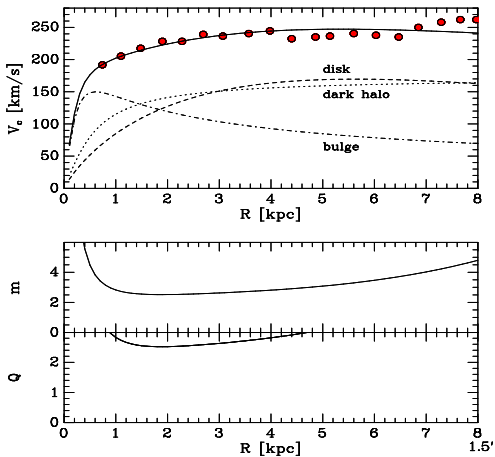}
\caption{Same as Fig.~\ref{fig9}, but for a submaximal disk component.}
\label{fig10}
\end{figure}

In Fig.~\ref{fig10} I show a submaximal disk model for the rotation curve of 
NGC\,2985. The fit to the rotation curve, which includes now a dark matter halo 
contributing as much as the disk, is of the same quality as in the case of a 
maximum disk, but both the predicted number of spiral arms and the $Q$ 
parameters as well as the implied mass-to-light ratio indicate that this not
viable model for the disk of NGC\,2985.

\section{Example: Diagnostics of the LSB galaxy F568-1}
In Fuchs (2002, 2003) I have applied similar diagnostics as in the previous 
section to a number of low surface brightness galaxies and describe here a 
typical result for the example of F568-1.
\begin{figure}
\includegraphics[width=40mm,height=40mm]{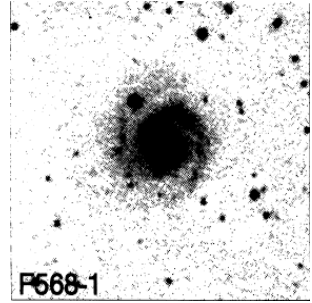}
\caption{Image of F568-1.}
\label{fig11}
\end{figure} 
The image of the galaxy, reproduced from \cite{Blo95}
in Fig.~\ref{fig11}, shows a regular two-armed spiral pattern. The rotation 
curve of F568-1 has been observed by \cite{Blo01}. Unfortunately the velocity
dispersions of stars in LSB disks have not been measured up to now. 
De Blok et al.~(2001) have provided -- as for each galaxy in their sample -- 
four types of models for the rotation curve of F568-1. The first model considers
only a dark halo, the next includes the contribution by the interstellar gas,
and the third model adds the contribution by a stellar disk with a, according
to population synthesis, `reasonable' mass-to-light ratio of 
$M/L_R$ = 1.4 $M_\odot/L_\odot$, which turns out to be a submaximal disk model.
The fourth model is a maximum disk model. This suite of models is presented for
the two cases that the dark halo is either a quasi-isothermal sphere or has a 
NFW density profile. Within each suite the reduced $\chi^2$s are practically 
indistinguishable from each other. The surprising result is, when I constrain 
the decomposition of the rotation curve dynamically using the density wave 
argument discussed in section (4), I find a mass-to-light ratio of 
$M/L_R$ = 14! 

\cite{Lee04} have attempted to modify the population synthesis model for LSB
galaxies toaccommodate such high mass-to-light ratios. They treated the stellar
population as an aging star burst with a bottom heavy IMF $\xi \sim
{\mathcal{M}}^{-(1+\alpha)}$ (with the notation that $\alpha$ = 1.8 corresponds
to a Salpeter law). Moreover, they varied the metallicities of the stars in
the range from [Fe/H] = -2.5 to 0.4. They find that they can indeed reproduce
the mass-to-light ratios and the $(B-R)_0$ colours of about 0.8 mag, which are
typical for the LSB galaxies in the sample of \cite{Fu03}, by adopting ages of 
the disks of 3 to 5 Gyrs, metallicities of [Fe/H] = -1.5 to -1, and an index 
of the IMF as steep as 
$\alpha$ = 2.35 to 2.85, respectively. These parameters were chosen rather 
ad hoc to explain the results of \cite{Fu03}. However, recently Kroupa and 
collaborators have developed the concept of an `integrated galactic initial 
mass function' which predicts very steep IMF indices $\alpha$. For instance, 
\cite{WeiKrou05} predict for galactic disks with masses of several
10$^{10}$ $M_\odot$, which I find typically for the LSBs in my sample,
in their `maximal' model an $\alpha$ = 2.6 (in the notation of Lee et al.~2004).
The IGMF concept has not yet been adapted in detail to LSBs, but looks very 
promising and we hope to address this in the near future.

\section{Conclusions}
Imaging and spectroscopy of galactic disks allows to infer their dynamics.
However, the decomposition of the rotation curves of spiral galaxies is highly
degenerate and needs further constraints. These can be found by considering the
implications for the dynamics of the disks of the galaxies. A useful diagnostic 
of the dynamics of a galactic disk is the Toomre $Q$ stability parameter. 
Another diagnostic is provided by the density wave theory of galactic spiral 
arms which predicts the number of spiral arms. Both diagnostics can be used to
constrain the masses of galactic disks. As an example of a high surface 
brightness spiral I have analyzed NGC\,2985 along these lines and found that it
seems to be dominated in the inner part, where the optical disk resides, 
by baryons. The resulting mass-to-light ratio is consistent with population
synthesis models for such galaxies. The dynamical analysis of the disks of low 
surface brightness galaxies, on the other hand, implies unexpected high
mass-to-light ratios. Such mass-to-light ratios can be explained by population
synthesis models with a bottom heavy IMF.

\newpage


\begin{thebibliography}{} 
  \bibitem[van Albada et al.~(1985)]{vanA85} van Albada, T.S., Bahcall, J.N.,
  Begeman, K., Sancisi, R.: 1985, ApJ 295, 305
  \bibitem[Arifyanto \& Fuchs (2001)]{ArFu01} Arifyanto, M.I., Fuchs, B.: 2001, 
  AGM 18, 160
  \bibitem[Athanassoula et al.~(1987)]{Ath87} Athanassoula, E., Bosma, A.,
  Papaioannou, S.: 1987, A\&A 179, 23
  \bibitem[Bell \& de Jong (2001)]{BeJo01} Bell, E.F., de Jong, R.S.: 2001, 
  ApJ 550, 212
  \bibitem[de Blok et al.~(1995)]{Blo95} de Blok, W.J.G., van der Hulst, 
  J.M., Bothun, G.D.: 1995, MNRAS 274, 235
  \bibitem[de Blok et al.~(2001)]{Blo01} de Blok, W.J.G., McGaugh, S.M., 
  Rubin, V.C.: 2001, AJ 122, 2396
  \bibitem[Bottema (1993)]{Bot93} Bottema, R.: 1993, A\&A 275, 16 
  \bibitem[Dubinski et al.~(2008)]{Dub08} Dubinski, J., Gauthier, J.-R.,
  Widrow, L., Nickerson, S.: 2008, in: J.G Funes, E.M Corsini (eds.)
  Formation and Evolution of Galaxy Disks (arXiv:0802.3997)
  \bibitem[Freeman's (1970)]{Fre70} Freeman, K.C.: 1970, ApJ 160, 811
  \bibitem[Fuchs (1999)]{Fu99} Fuchs, B.: 1999, ASP Conf. Ser. 182, 365 
  \bibitem[Fuchs (2001)]{Fu01} Fuchs, B.: 2001, A\&A 368, 107
  \bibitem[Fuchs (2002)]{Fu02} Fuchs, B.: 2002, in: H.V. Klapdor-Kleingrothaus,
  	R.D. Viollier (eds.) Dark matter in astro- and particle physics,
	Springer, Berlin, p. 28 
  \bibitem[Fuchs (2003)]{Fu03} Fuchs, B.: 2003, Ap\&SS 284, 719
  \bibitem[Fuchs \& v.~Linden (1998)]{FuLi98} Fuchs, B., von Linden, S.: 1998,
  MNRAS 294, 513
  \bibitem[Gerssen et al.~(1997)]{Ge97} Gerssen, J., Kuijken, K., Merrifield,
  M.R.: 1997, MNRAS 288, 618
  \bibitem[Gerssen et al.~(2000)]{Ge00} Gerssen, J., Kuijken, K., Merrifield,
  M.R.: 2000, MNRAS 317, 545
  \bibitem[Hockney \& Hohl (1969)]{Ho69} Hockney, R.W., Hohl, F.: 1969, 
  AJ 74, 1102
  \bibitem[Lee et al.(2004)]{Lee04} Lee, H.-y., Gibson, B.K., Flynn, C., Kawata,
  D. Beasley, M.A.: 2004, MNRAS 353, 113 
  \bibitem[(Lindblad 1959)]{Li59} Lindblad, B.: 1959, 
  {\it Handbuch der Physik} 53, 21 
  \bibitem[(Navarro et al.~1997)]{Nav97} Navarro, J.F. Frenk, C.S., White,
  S.D.M.: 1997, ApJ 490, 493
  \bibitem[Prugniel et al.~(1998)]{Pru98} Prugniel, Ph., Zasov, A., Busarello,
  G., Simien, F.: 1998, A\&AS 127, 117
  \bibitem[Rubin et al.~(1978)]{RuFo78} 
  Rubin, V.C., Ford, W.K., Thonnard, N.: 1978, ApJ 225, L107
  \bibitem[Toomre (1964)]{Too64} Toomre, A.: 1964, ApJ 139, 1217
  \bibitem[Toomre (1981)]{Too81} Toomre, A.: 1981, in: S.M. Fall, 
  D. Lynden--Bell (eds.) The Structure and Evolution of Normal Galaxies,
  Cambridge Univ. Pres, Cambridge, p. 111
  \bibitem[(Verheijen et al.~2004)]{Ver04} Verheijen, M.A.W., Bershady, M.A.,
  Andersen, D.R. et al.: 2004, AN 325, 151
  \bibitem[Weidner \& Kroupa (2005)]{WeiKrou05} Weidner, C., Kroupa, P.: 2005,
  ApJ 625, 754
\end{thebibliography}
\end{document}